\documentclass[aps,showpacs,11pt]{revtex4}
\usepackage{dcolumn}
\usepackage{graphicx}
\usepackage{amsmath}
\usepackage{amsfonts}
\usepackage{amssymb}
\usepackage{psfrag}
\usepackage{wrapfig}
\usepackage{subfigure}
\usepackage{makeidx}
\usepackage{bm}
\usepackage{epsf}
\usepackage{hyperref}
\makeatletter

\newcommand{\Rmnum}[1]{\expandafter\@slowromancap\romannumeral #1@}
\makeatother

\begin{document}

\title{Critical behavior of charged Gauss-Bonnet-AdS black holes in the grand canonical ensemble}

\author{De-Cheng Zou}
\email{zoudecheng@sjtu.edu.cn}
\author{Yunqi Liu}
\email{liuyunqi@sjtu.edu.cn}
\author{Bin Wang}
\email{wang$_$b@sjtu.edu.cn}

\affiliation{Department of Physics and Astronomy, Shanghai Jiao Tong University, Shanghai 200240, China}

\date{\today}

\begin{abstract}
\indent

We study the thermodynamics in the grand
canonical ensemble of D-dimensional charged
Gauss-Bonnet-AdS black holes in the extended
phase space. We find that the usual small-large
black hole phase transition, which exhibits
analogy with the Van de Waals liquid-gas system
holds in five-dimensional spherical charged
Gauss-Bonnet-AdS black holes when its potential
is fixed within the range
$0<\Phi<\frac{\sqrt{3}\pi}{4}$. For the other
higher dimensional and topological charged
Gauss-Bonnet-AdS black holes, there is no such
phase transition. In the limiting case,
Reissner-Nordstrom-AdS black holes, with
vanishing Gauss-Bonnet parameter, there is no
critical behavior in the grand canonical
ensemble. This result holds independent of the
spacetime dimensions and topologies. We also
examine the behavior of physical quantities in
the vicinity of the critical point in the
five-dimensional spherical charged
Gauss-Bonnet-AdS black holes.

\end{abstract}



\maketitle
\section{Introduction}
\label{1s}

Black hole thermodynamics has been an intriguing
subject of discussions for decades. In view of
AdS/CFT correspondence, the black hole
thermodynamics in the presence of a negative
cosmological constant becomes more interesting
nowadays. The thermodynamic properties of AdS
black holes was initially studied in
\cite{Hawking:1982dh}, where a Hawking-Page phase
transition between the phase spaces of the
Schwarzschild AdS black hole and pure AdS space
was disclosed. Later, more interesting
discoveries were obtained for the charged AdS
black holes \cite{Chamblin:1999tk,
Chamblin:1999hg}, where it was found that a first
order small black hole and big black hole phase
transition is allowed in the canonical ensemble
where the black hole charge is kept fixed. This
phase transition is superficially analogous to a
liquid-gas phase transition of the Van der Waals
fluid. This superficial reminiscence was also
observed in other AdS backgrounds
\cite{Niu:2011tb,Dey:2006ds,
Dey:2007vt,Anninos:2008sj,Fernando:2006gh,Poshteh:2013pba,Mo:2013sxa,Lala:2012jp,Wei:2012ui,
Tsai:2011gv,Banerjee:2011au,Banerjee:2012zm,Banerjee:2010bx,Banerjee:2010qk,Banerjee:2010da}.
The understanding of the phase transition and the
critical phenomena has been further extended to
more complicated background, such as the charged
Gauss-Bonnet(GB)-AdS black holes
\cite{Dey:2007vt,Dey:2006ds}.  In the canonical
ensemble where the black hole charge is fixed,
the phase transition between small and big black
holes exits as well in the charged GB-AdS holes.
The discussion has also been generalized to the
grand canonical ensemble where the black hole is
allowed to emit and absorb charged particles
keeping the potential fixed till the thermal
equilibrium is reached. For the spherical
Reissner-Nordstrom(RN)-AdS black hole, there is
no critical behavior observed in the grand
canonical ensemble \cite{Chamblin:1999tk,
Chamblin:1999hg}. While for the
spherical charged GB-AdS black hole background in
five-dimensions, the phase transition between the
small and large black holes in the grand
canonical ensemble can still happen
\cite{Dey:2007vt,Anninos:2008sj}.

In the usual discussions of thermodynamical properties of black holes in (A)dS spaces,
the cosmological constant is treated as a fixed parameter.
Recently the study of thermodynamics in AdS black
holes has been generalized to the extended phase
space, where the cosmological constant is regarded
as a variable and also identified with thermodynamic pressure
\cite{Dolan:2011xt,Dolan:2010ha}
\begin{eqnarray}
P=-\frac{\Lambda}{8\pi}=\frac{(D-1)(D-2)}{16\pi l^2}\label{eq:1a}
\end{eqnarray}
in the geometric units $G_N=\hbar=c=k=1$.  Here
$D$ stands for the number of spacetime dimensions
and $l$ denotes the AdS radius.
Taking the cosmological constant as a thermodynamic
pressure is nowadays a common practice, where such operations implicitly assume
that gravitational theories including different values of the
cosmological constants fall in the ``same class",
with unified thermodynamic relations. The
common excuse for doing this is that the classical
theory of gravity may be an effective
theory which follows from a yet unknown fundamental theory,
in which all the presently ``physical constants" are actually
moduli parameters that can run from place to place
in the moduli space of the fundamental theory.
Since the fundamental theory is yet unknown,
it is more reasonable to consider the extended thermodynamics of gravitational
theories involving only a single action,
and then all variables will appear in the thermodynamical relations.
Similar situations the parameter $b$ of Born-Infeld term \cite{Gunasekaran:2012dq}
and coupling coefficient of GB term \cite{Cai:2013qga} can be evaluated
in different gravitational theories.
In addition, the variation of the cosmological constant is
included in the first law of black hole thermodynamics, which
ensures the consistency between the first law of
black hole thermodynamics  and the Smarr formula.
Including the variation of
the cosmological constant in the first law, the
AdS black hole mass is identified with enthalpy
and there exists a natural conjugate
thermodynamic volume to the cosmological
constant. In the extended phase space with
cosmological constant and volume as thermodynamic
variables, it was interestingly observed that the
system admits a more direct and precise
coincidence between the first order small-large
black hole phase transition and the liquid-gas
change of phase occurring in fluids
\cite{Kubiznak:2012wp}.  More discussions on
phase transitions in AdS black holes by treating
the cosmological constant as a dynamical quantity
can be found in
\cite{Gunasekaran:2012dq,Hendi:2012um,Belhaj:2012bg,
Belhaj:2013ioa, Chen:2013ce,
Zhao:2013oza,Belhaj:2013cva, Xu:2013zea,
Dutta:2013dca, Zou:2013owa, Altamirano:2013ane,
Altamirano:2013uqa, Cai:2013qga,Altamirano:2014tva,
Ma:2013aqa,Mo:2014qsa,Wei:2014hba,Mo:2013ela,Mo:2014mba,
Zhang:2014jfa,Mo:2014lza}.
However, all these discussions were concentrated
on the canonical ensemble by fixing the charge of
the black hole in the extended phase space.

It is of great interest to generalize the
discussion of the phase transitions of black
holes in the extended phase space  to the grand
canonical ensemble where the background AdS black
hole has a constant fixed potential.   In the
footnote of \cite{Kubiznak:2012wp}, it was
briefly argued that in the extended phase space
the criticality cannot happen in the grand
canonical ensemble for the four-dimensional
spherical RN-AdS black hole.  It is interesting
to ask whether in the extended phase space the
critical behavior of the RN-AdS black hole will
emerge for different numbers of spacetime
dimensions or other topologies. Whether in the
grand canonical ensemble,  the critical behavior
in the extended phase space can be in consistent
with the result when the cosmological constant
was fixed \cite{Chamblin:1999tk}. This is the
first motivation of the present paper. Besides,
we would like to generalize the exploration of
the critical behaviors to a more complicated
black hole background, the charged GB-AdS black
hole, in the grand canonical ensemble.
Until now, $P-V$ criticality of
GB-AdS black hole has been discussed in the canonical
ensemble \cite{Cai:2013qga}, in which the coupling coefficient of GB term
has been regarded as a variable appearing in the first
law of black hole thermodynamics. On the other hand,
in the GB gravity, the coupling coefficient of GB term
should be also regarded as a variable in order to get a consistent
Smarr relation for black hole thermodynamics, which has been
verified in \cite{Kastor:2010gq} by employing the Hamiltonian
perturbation theory techniques. Moreover,
similar situation occurred for the Born-Infeld black holes
\cite{Breton:2004qa,Gunasekaran:2012dq,Rasheed:1997ns}.
In this paper, we will concentrate on the extended phase space with
fixed potential and examine whether  the critical
behavior in the charged GB-AdS black holes can
happen once the black hole pressure and volume
are identified. We will disclose how the
criticality will be influenced by spacetime
dimension and topology in the higher order derivative
gravity.

This paper is organized as follows. In
Sec.~\ref{2s}, we will firstly present the
solutions of the $D$-dimensional charged GB-AdS
black holes. Then  we will examine the critical
behaviors in the charged GB-AdS black holes in
the grand canonical ensemble for fixed potential.
In Sec.~\ref{3s}, we will discuss the influence of
the critical behavior by topologies and spacetime
dimensions. Later in Sec.~\ref{4s} we will study
the critical point by using  the Ehrenfest
equations. Finally in Sec.~\ref{5s} we will
present our closing remarks.

\section{Charged Gauss-Bonnet AdS black hole background }
\label{2s}

The solution of $D$-dimensional charged GB-AdS black hole with negative cosmological constant
is  \cite{Cai:2013qga}
\begin{eqnarray}
&&ds^2=-f(r)dt^2+\frac{1}{f(r)}dr^2+r^2h_{ij}dx^idx^j,\label{eq:2a}\\
f(r)&=&k+\frac{r^2}{2\tilde{\alpha}_{GB}}\left[1-\sqrt{1-\frac{64\pi\tilde{\alpha}_{GB}P}{(D-1)(D-2)}
+\frac{64\tilde{\alpha}_{GB}\pi M}{(D-2)\Sigma_k r^{D-1}}
-\frac{2\tilde{\alpha}_{GB}Q^2}{(D-2)(D-3)r^{2D-4}}}\right],\label{eq:3a}
\end{eqnarray}
where $\Sigma_k$ is the volume of the $(D-2)$-dimensional unit hypersurface,
$\tilde{\alpha}_{GB}=(D-3)(D-4)\alpha_{GB}$, $M$ is the black hole mass and $Q$ is related to
the charge of the black hole. We will only consider the
positive GB coefficient $\alpha_{GB}$ in the following discussion.
The coordinates are labeled as $x^{\mu}=(t, r, x^i)$, $(i=1,\cdots ,D-2)$ and the metric function
$h_{ij}$ is a function of the coordinates $x^i$, which span an $(D-2)$-dimensional hypersurface with
constant scalar curvature $(D-2)(D-3)k$. The constant $k$ characterizes the geometric property of
hypersurface, which takes values $k=0$  for flat, $k=-1$ for
negative curvature and $k=1$ for positive curvature, respectively.
If we take $\tilde{\alpha}_{GB}\rightarrow 0$, the solution $f(r)$ reduces to the RN-AdS case.
In order to have a well-defined vacuum solution with
$M=Q=0$, the Gauss-Bonnet coefficient $\tilde{\alpha}_{GB}$ and pressure $P$ have to satisfy the
following constraint
\begin{eqnarray}
0<\frac{64\pi\tilde{\alpha}_{GB}P}{(D-1)(D-2)}<1.\label{eq:6a}
\end{eqnarray}

In terms of the horizon radius $r_+$, the mass $M$, Hawking temperature $T$,
entropy $S$ and electromagnetic potential $\Phi$ of charged GB-AdS black holes
can be written as \cite{Cai:2013qga}
\begin{eqnarray}
M&=&\frac{(D-2)\Sigma_kr_+^{D-3}}{16\pi}\left[k+\frac{k^2\tilde{\alpha}_{GB}}{r_+^2}
+\frac{16\pi P r_+^2}{(D-1)(D-2)}+\frac{2Q^2}{(D-2)(D-3)r_+^{2D-6}}\right],\label{eq:4a}\\
T&=&\frac{1}{4\pi r_+\left(r_+^2+2k\tilde{\alpha}_{GB}\right)}\left[\frac{16\pi
Pr_+^4}{D-2}+(D-3)k r_+^2
+(D-5)k^2\tilde{\alpha}_{GB}-\frac{2 Q^2}{(D-2)r_+^{2D-8}}\right] \label{eq:5a},\\
S&=&\int^{r_+}_{0}{T^{-1}\frac{dM}{dr_+}dr_+}=\frac{\Sigma_kr_+^{D-2}}{4}\left[1+\frac{2k(D
-2)\tilde{\alpha}_{GB}}{(D-4)r_+^2}\right],\quad \Phi=\frac{\Sigma_k Q}{4\pi(D-3) r_+^{D-3}}.\label{eq:7a}
\end{eqnarray}
Note that the entropy $S$ will take negative values
for sufficiently small black holes for hyperbolic charged GB-AdS black hole.
However, an ambiguity can be added into the definition of the entropy which
can be appropriately chosen to avoid the appearance of
negative entropies \cite{Clunan:2004tb,Nojiri:2002qn}.
Here we do not rewrite the expression for entropy $S$ with $k=-1$,
since the entropy of hyperbolic GB-AdS black holes will be not considered
in the next sections.

The expressions for mass $M$ and Hawking temperature $T$
of charged GB-AdS black holes with fixed potential can be rewritten as
\begin{eqnarray}
T&=&\frac{1}{4\pi r_+\left(r_+^2+2k\tilde{\alpha}_{GB}\right)}\left[\frac{16\pi Pr_+^4}{D-2}+(D-3)k r_+^2
+(D-5)k^2\tilde{\alpha}_{GB}-\frac{32\pi^2(D-3)^2\Phi^2r_+^2}{(D-2)\Sigma_k^2}\right],\label{eq:9a}\\
M&=&\frac{(D-2)\Sigma_kr_+^{D-3}}{16\pi}\left[k+\frac{k^2\tilde{\alpha}_{GB}}{r_+^2}
+\frac{16\pi P r_+^2}{(D-1)(D-2)}+\frac{32\pi^2(D-3)\Phi^2}{(D-2)\Sigma_k^2}\right].\label{eq:10a}
\end{eqnarray}

The black hole mass $M$ is considered as the
enthalpy rather than the internal energy of the
gravitational system \cite{Kastor:2009wy}.
Moreover, the thermodynamic quantities satisfy
the following differential relation
\begin{eqnarray}
dM=TdS+Qd\Phi+VdP+\Omega d\tilde{\alpha}_{GB},\label{eq:11a}
\end{eqnarray}
where $V$ denotes the thermodynamic volume with
\begin{eqnarray}
V=\left(\frac{\partial M}{\partial P}\right)_{S,\Phi,\tilde{\alpha}_{GB}}=\frac{\Sigma_k r_+^{D-1}}{D-1},
\quad \Omega=\left(\frac{\partial M}{\partial\tilde{\alpha}_{GB}}\right)_{S,\Phi,P}
=\frac{(D-2)k^2\Sigma_kr_+^{D-5}}{16\pi}.\label{eq:12a}
\end{eqnarray}
By the scaling argument, we can obtain the generalized Smarr relation for the charged GB-AdS black holes
in the extended phase space
\begin{eqnarray}
M=\frac{D-2}{D-3}TS+\Phi Q-\frac{2}{D-3}VP+\frac{2}{D-3}\Omega\tilde{\alpha}_{GB}. \label{eq:13a}
\end{eqnarray}

\section{Critical behaviors of charged GB-AdS black holes in the grand canonical ensemble}
\label{3s}

\textbf{
From Eq.~(\ref{eq:9a}), the equation of state $P(V,T,\tilde{\alpha}_{GB},\Phi)$ can be expressed into}
\begin{eqnarray}
P=\frac{(D-2)T}{4r_+}\left(1+\frac{2k\tilde{\alpha}_{GB}}{r_+^2}\right)-\frac{(D-2)(D-3)k}{16\pi r_+^2}
-\frac{(D-2)(D-5)k^2\tilde{\alpha}_{GB}}{16\pi r_+^4}+\frac{2\pi(D-3)^2\Phi^2}{\Sigma_k^2 r_+^2}.\label{eq:14a}
\end{eqnarray}
To compare with the Van der Waals fluid equation in $D$-dimensions, we can translate the ``geometric" equation
of state to physical one by identifying the specific volume $v$ of the fluid with
the horizon radius of the black hole as $v=\frac{4r_+}{D-2}$.

We know that the critical points occur when $P$
has an inflection point,
\begin{eqnarray}
\frac{\partial P}{\partial r_+}\Big|_{T=T_c, r_+=r_c}
=\frac{\partial^2 P}{\partial r_+^2}\Big|_{T=T_c, r_+=r_c}=0.\label{eq:15a}
\end{eqnarray}
Then we can obtain the critical temperature
\begin{eqnarray}
T_c=\frac{(D-3)kr_c}{2\pi\left(r_c^2+6k\tilde{\alpha}_{GB}\right)}
+\frac{(D-5)k^2\tilde{\alpha}_{GB}}{\pi\left(r_c^2
+6k\tilde{\alpha}_{GB}\right)r_c }-\frac{16\pi(D-3)^2\Phi^2r_c}{(D
-2)\left(r_c^2+6k\tilde{\alpha}_{GB}\right)\Sigma_k^2} \label{eq:16a}
\end{eqnarray}
and the equation for the critical horizon radius $r_c$ (specific volume $v_c=\frac{4r_c}{D-2}$) is
\begin{eqnarray}
\frac{8\pi (D-3)^2\left(r_c^2-6k\tilde{\alpha}_{GB}\right)r_c\Phi^2}{(D
-2)\left(r_c^2+6k\tilde{\alpha}_{GB}\right)\left(r_c^2
+12k\tilde{\alpha}_{GB}\right)\Sigma_k^2}-\frac{(D-3)kr_c^4
-12k^2\tilde{\alpha}_{GB}r_c^2+12k(D-5)\tilde{\alpha}_{GB}^2}{4\pi r_c\left(r_c^2
+6k\tilde{\alpha}_{GB}\right)\left(r_c^2+12k\tilde{\alpha}_{GB}\right)}=0,\label{eq:17a}
\end{eqnarray}
where $r_c$ denotes the critical value of $r_+$.

Taking $\tilde{\alpha}_{GB}=0$, one can reduce Eq.~(\ref{eq:17a}) to the RN-AdS case
\begin{eqnarray}
\frac{1}{r_c}\left(\frac{8\pi (D-3)\Phi^2}{(D-2)\Sigma_k^2}
-\frac{k}{4\pi}\right)=0.\label{eq:18a}
\end{eqnarray}
Different from the discussion in the canonical
ensemble \cite{Gunasekaran:2012dq}, $r_c$
maintains the same order in the two terms related
to $\Phi$ and $k$ in Eq.~(\ref{eq:18a}). For $k=
0$, Eq.~(\ref{eq:18a}) cannot be satisfied for
nonzero $\Phi$. For $k=-1$, the right hand side
of Eq.~(\ref{eq:18a}) can never be zero. For
$k=1$, Eq.~(\ref{eq:18a}) can hold only when
$\Phi=\frac{\Sigma_k}{4\pi}\sqrt{\frac{(D-2)}{2(D-3)}}$.
However, when $k=1$ and
$\Phi=\frac{\Sigma_k}{4\pi}\sqrt{\frac{(D-2)}{2(D-3)}}$,
both the critical pressure $P_c$ and the critical
temperature $T_c$ vanish. This result tells us
that in the grand canonical ensemble, there is no
critical behavior between small-large black holes
in the RN-AdS backgrounds in the extended phase
space. This result holds independent of the
spacetime dimensions and topology.  This
generalized the first observation in the
four-dimensional spherical RN-AdS black holes
with fixed potential in the extended phase space
\cite{Kubiznak:2012wp}. The obtained result in
the extended phase space supports the arguments
with fixed cosmological constant
\cite{Chamblin:1999tk}.

Now we would like to extend the discussion of the
critical behavior to the charged GB-AdS black
holes in the extended phase space with fixed
potential.  Looking at the critical temperature
$T_c$ (Eq.~(\ref{eq:16a})), we find that it is
negative for $k=0$, which indicates that no
criticality can happen when $k=0$.  Thus we only
need to focus on the cases with  $k=-1$ and
$k=1$.

\subsection{Five-dimensional Charged GB-AdS black holes with fixed potential}

We first concentrate on the five dimensional
spacetime, $D=5$, so that Eq.~(\ref{eq:17a})
reduces to
\begin{eqnarray}
\frac{\left(64\pi^2\Phi^2-3k\Sigma_k^2\right)\left(r_c^2
-6k\tilde{\alpha}_{GB}\right)r_c}{6\pi\Sigma_k^2\left(r_c^2
+6k\tilde{\alpha}_{GB}\right)\left(r_c^2+12k\tilde{\alpha}_{GB}\right)}=0.\label{eq:19a}
\end{eqnarray}
For $k=-1$, it is impossible to have a physical
solution of $r_c$ to make the above equation to
be satisfied. This corresponds to say that there
does not exist the Van der Waals like small-large
black hole phase transition in the hyperbolic
space.

For $k=1$, $\Sigma_k$ equals to $2\pi^2$.
When $3\pi^2-16\Phi^2=0$, namely $\Phi=\frac{\sqrt{3}\pi}{4}$, $r_c$ can take any positive
values, while the critical temperature $T_c$ and $P_c$ both disappear from Eqs.(15)(17).
In case of $\Phi\neq\frac{\sqrt{3}\pi}{4}$, the solution of Eq.~(20) and corresponding
critical values of temperature and pressure read
\begin{eqnarray}
r_c=\sqrt{6\tilde{\alpha}_{GB}}, \quad T_c=\frac{3\pi^2
-16\Phi^2}{6\sqrt{6\tilde{\alpha}_{GB}}\pi^3},
\quad P_c=\frac{3\pi^2-16\Phi^2}{144\pi^3\tilde{\alpha}_{GB}}.\label{eq:20a}
\end{eqnarray}
It is easy to see that for the potential with
fixed values in the range
$0<\Phi<\frac{\sqrt{3}\pi}{4}$, criticality can
appear in this charged GB-AdS black hole
background. This result holds for all positive GB
coupling constants $\tilde{\alpha}_{GB}$.
Considering the specific volume
$v_c=\frac{4r_c}{D-2}$, we can obtain an
interesting relation
$\frac{P_cv_c}{T_c}=\frac{1}{D-2}=\frac{1}{3}$
from Eq.~(\ref{eq:20a}), which is
independent of $\tilde{\alpha}_{GB}$ and potential $\Phi$.
We plot the $P-r_+$ isotherm diagram with
different values of $\Phi$ for $D=5$ in Fig.~1.
It shows that for $\Phi=1<\frac{\sqrt{3}\pi}{4}$,
the dashed line corresponds to the ``idea gas"
phase behavior when $T>T_c$, and the Van der Waals
like small-large black hole phase transition
appears in the system when $T<T_c$.

\begin{figure}[h]
\centering
\includegraphics{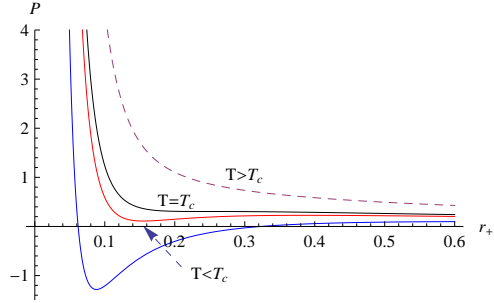}
\caption{ The $P-r_+$ diagram of spherical GB-AdS black holes in
the grand canonical ensemble for $\Phi=1$, $\tilde{\alpha}_{GB}=0.01$ and $D=5$.}
\end{figure}

On the other hand, in the Van der Waals fluid we have the liquid-gas
phase transition equation
\cite{Goldenfeld:1992qy}
\begin{eqnarray}
\left(P+\frac{a}{v^2}\right)\left(v-b\right)=K T,\label{eq:21a}
\end{eqnarray}
where $v$ is the specific volume of the fluid,
$P$ is the pressure, $T$ is the temperature, and
$K$ is the Boltzmann constant. The constant $b>0$
describes the molecules with nonzero size in the
fluid, and the constant $a>0$ is a measure of the
attraction between them. Defining
\begin{eqnarray}
p=\frac{P}{P_c}, \quad \nu=\frac{v}{v_c}, \quad \tau=\frac{T}{T_c}, \label{eq:21b}
\end{eqnarray}
the Van der Waals fluid equation can be rewritten
in a dimensionless form
\begin{eqnarray}
8\tau=\left(3\nu-1\right)\left(p+\frac{3}{\nu^2}\right).\label{eq:21c}
\end{eqnarray}
Here the compressibility factor $\frac{P_cv_c}{KT_c}=\frac{3}{8}$
is a universal number for all fluids.
For the $D$-dimensional RN-AdS black hole, the analogy of the Van
der Waals liquid-gas phase transition equation is
described by \cite{Gunasekaran:2012dq}
\begin{eqnarray}
4(D-2)\tau=(2D-5)\nu\left(p+\frac{D-2}{(D-3)\nu^2}\right)-\frac{1}{(D-3)\nu^{2D-5}}.\label{eq:21f}
\end{eqnarray}
The ratio is
$\frac{P_cv_c}{T_c}=\frac{2D-5}{4D-8}$ \cite{Gunasekaran:2012dq, Belhaj:2012bg},
which depends on the spacetime dimensions.
For $D=5$, Eq.~(\ref{eq:21f}) reduces to
\begin{eqnarray}
12\tau=5\nu\left(p+\frac{3}{2\nu^2}\right)-\frac{1}{2\nu^5}\label{eq:21d}
\end{eqnarray}
with the ratio $\frac{P_cv_c}{T_c}=5/12$.
For the five dimensional charged GB-AdS black hole with fixed potential,
Eq.~(\ref{eq:14a}), the analogy of the Van der Waals liquid-gas phase transition equation
has the form
\begin{eqnarray}
\tau=\frac{3\nu\left(3+p\nu^2\right)}{4\left(2+3\nu^2\right)}.\label{eq:21e}
\end{eqnarray}
One can see the Van der Waals like small-large black hole
phase transition happen for the these black hole backgrounds.
However, the analogy of the Van der Waals liquid-gas phase transition equations
can take different forms, and the ratio $\frac{P_cv_c}{T_c}$ arrives at different values
for various black hole backgrounds.

The behavior of the Gibbs free energy $G$ is
important to determine the thermodynamic phase
transition. In the grand canonical ensemble with
fixed potential, the Gibbs free energy $G$ obeys
the following thermodynamic relation
\begin{eqnarray}
G=M-TS-Q\Phi&=&\frac{1}{24\pi\left(r_+^2+2\tilde{\alpha}_{GB}\right)}\left[3\pi^2(r_+^4
-3\tilde{\alpha}_{GB}r_+^2+6\tilde{\alpha}_{GB}^2)\right.\nonumber\\
&&\left.-4\pi^3r_+^4(r_+^2+18\tilde{\alpha}_{GB})P-16r_+^2(r_+^2
-6\tilde{\alpha}_{GB})\Phi^2\right].\label{eq:22a}
\end{eqnarray}
Here $r_+$ is understood as a function of
pressure and temperature, $r_+=r_+(P,T)$, via
equation of state Eq.~(\ref{eq:14a}). In Fig.~2,
we see that the $G$ surface demonstrates the
characteristic ``swallow tail" behavior, which
shows that there is a Van der Waals like first
order phase transition in the system. We also
plot the coexistence line in the $(P,T)$ plane by
finding a curve where the Gibbs free energy and
temperature coincide for small and large black
holes as shown in Fig.~2. The coexistence line in
the $(P,T)$ plane is very similar to that in the
Van der Waals fliud. The critical point is shown
by a small circle at the end of the coexistence
line. The small-large black hole phase transition
occurs for $T<T_c$.

\begin{figure}[h]
\centering
\includegraphics{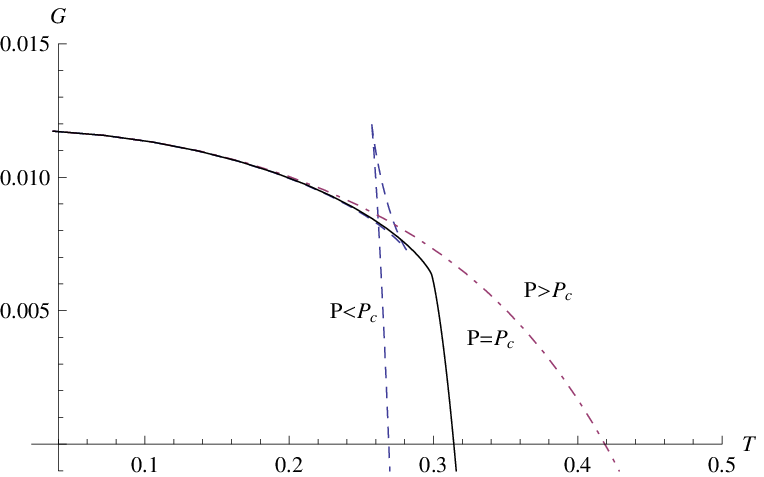}
\hfill%
\includegraphics{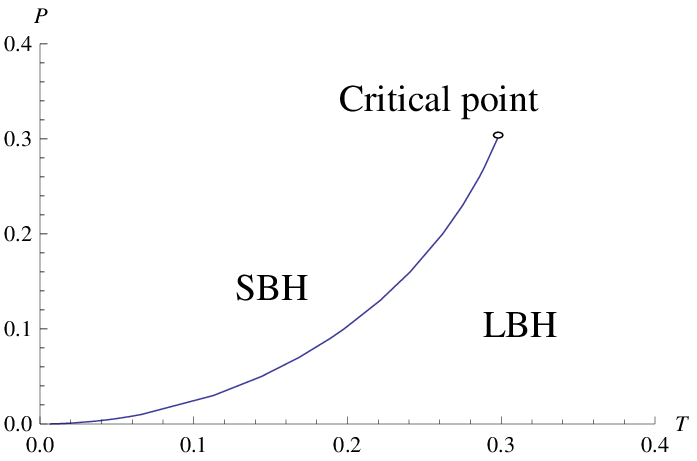}
\caption{ The Gibbs free energy $G$ of spherical
charged GB-AdS black holes and coexistence line
of small-large black holes  phase transition in
the grand canonical ensemble for $\Phi=1$,
$\tilde{\alpha}_{GB}=0.01$ and $D=5$.}
\end{figure}

\subsection{Higher dimensional charged GB-AdS black holes with Fixed potential}

Defining $\tilde{\Phi}=\frac{\pi\Phi}{\Sigma_k}$,
equation Eq.~(\ref{eq:17a}) can be rewritten as
\begin{eqnarray}
&&\frac{1}{4(D-2)\pi r_c(r_c^2+6k\tilde{\alpha}_{GB})(r_c^2+12k\tilde{\alpha}_{GB})}\left[(D-2)\left(-(D-3)kr_c^4
+12k^2r_c^2\tilde{\alpha}_{GB}-12k(D-5)\tilde{\alpha}_{GB}^2\right)\right.\nonumber\\
&&\left.+32(D-3)^2r_c^2(r_c^2-6k\tilde{\alpha}_{GB})\tilde{\Phi}^2\right]=0.\label{eq:23a}
\end{eqnarray}
For $k=-1$, the right hand side of
Eq.~(\ref{eq:23a}) can never disappear for any
real value of $r_c$. This tells us that there
does not exist any critical behavior in the
hyperbolic changed GB-AdS black holes.

Now we turn our discussion to the spherical case
with $k=1$.  The Eq.~(\ref{eq:23a}) reduces to
\begin{eqnarray}
&&(D-3)\left[32(D-3)\tilde{\Phi}^2-(D-2)\right]r_c^4-12\tilde{\alpha}_{GB}\left[16(D
-3)^2\tilde{\Phi}^2-(D-2)\right]r_c^2\nonumber\\
&&-12(D-5)(D-2)\tilde{\alpha}_{GB}^2=0.\label{eq:24a}
\end{eqnarray}
For $D\geq 6$, we have the relation
$\frac{(D-2)}{16(D-3)^2}<\frac{(D-2)^2}{48(D-3)^2}<\frac{(D-2)}{32(D-3)}$.
When the potential is fixed with the value
$\tilde{\Phi}^2=\frac{(D-2)}{32(D-3)}>\frac{(D-2)}{16(D-3)^2}$,
we cannot find any real solution of $r_c$ from
Eq.~(\ref{eq:24a}).  If the potential is fixed
with other values,
$\tilde{\Phi}^2\neq\frac{(D-2)}{32(D-3)}$, we can
obtain the solution of Eq.~(\ref{eq:24a}) in the
form
\begin{eqnarray}
r_{c\pm}^2=\frac{\tilde{\alpha}_{GB}\left[12\left(16(D-3)^2\tilde{\Phi}^2-(D-2)\right)\pm\sqrt{\Delta}\right]}{2(D
-3)\left[32(D-3)\tilde{\Phi}^2-(D-2)\right]},\label{eq:25a}
\end{eqnarray}
where
$\Delta=48\left[768(D-3)^4\tilde{\Phi}^4+32(D-2)(D-8)(D-3)^2\tilde{\Phi}^2-(D-6)(D-2)^3\right]$.
To keep the $\Delta$ non-negative, we need the
potential satisfy
$\tilde{\Phi}^2\geq\frac{(D-2)^2}{48(D-3)^2}$.
When the potential is fixed in the range
$\frac{(D-2)^2}{48(D-3)^2}\leq\tilde{\Phi}^2<\frac{(D-2)}{32(D-3)}$,
we have
$12\left(16(D-3)^2\tilde{\Phi}^2-(D-2)\right)>\sqrt{\Delta}\geq0$.
The numerator of Eq.~(\ref{eq:25a}) is always
positive independent of the sign we choose in
front of the square root. However the denominator
of Eq.~(\ref{eq:25a}) is always negative in this
fixed potential range. Thus in the above chosen
range of the potential, there is no criticality
to happen. We can also fix the potential in the
range $\tilde{\Phi}^2>\frac{(D-2)}{32(D-3)}$ to
avoid the negative $\Delta$.  Now the denominator
in Eq.~(\ref{eq:25a}) can always remain
positive. In the numerator when the potential is
fixed in this range,
$\sqrt{\Delta}>12\left(16(D-3)^2\tilde{\Phi}^2-(D-2)\right)>0$,
 we can only select the positive sign in
front of the square root term. We can always find
the physical solution of $r_c$ when the potential
is fixed in the range $\tilde{\Phi}^2>\frac{(D
-2)}{32(D-3)}$.

Can the criticality appear for the higher
dimensional spherical charged GB-AdS black holes?
To answer this question, let us further examine
the critical temperature $T_c$. From
Eq.~(\ref{eq:16a}) we have
\begin{eqnarray}
T_c=\frac{-(D-3)^{3/2}\left[32(D-3)\tilde{\Phi}^2-(D-2)\right]^{3/2}\left(\kappa+\sqrt{\Delta}/4\right)
}{6(D-2)\sqrt{2\tilde{\alpha}_{GB}}\pi\left[3\left(16(D-3)^2\tilde{\Phi}^2-(D-2)\right)
+\sqrt{\Delta}/4\right]^{1/2}\left(\kappa+\sqrt{\Delta}/12\right)},\label{eq:26a}
\end{eqnarray}
where $\kappa=48(D-3)^2\tilde{\Phi}^2-(D-2)^2$.
It is worth noting that  $\kappa>0$ when the
potential is fixed in the range
$\tilde{\Phi}^2>\frac{(D-2)}{32(D-3)}$, so that
the critical temperature $T_c$ is always negative
in this potential range. Therefore the
criticality  cannot appear and there is no
small-large black hole phase transition, in
resemblance with the Van der Waals phase
transition, for higher dimensional charged GB-AdS
black holes in the grand canonical ensemble.

In summary, the Van der Waals like phase
transition only happens in the $five-dimensional$
spherical charged GB-AdS black hole in the grand
canonical ensemble when the potential is fixed
within the range $0<\Phi<\frac{\sqrt{3}\pi}{4}$.
The criticality cannot appear in the spacetime
with other dimensions and topologies.  To further
examine the critical behavior in the
five-dimensional spherical charged GB-AdS black
holes, in the following we investigate the
behaviors of thermodynamic quantities near the
critical point of the phase transition.

\subsection{Critical exponents near critical point}

Now we turn to compute the critical exponents
$\alpha$, $\beta$, $\gamma$, $\delta$ for the
black hole system, which characterize the
behaviors of physical quantities in the vicinity
of the critical point $(r_+=r_c, v=v_c, T=T_c,
P=P_c)$ for the five-dimensional spherical
charged GB-AdS black hole in the grand canonical
ensemble. Near the critical point,  critical
exponents are defined as follows
\cite{Kubiznak:2012wp}
\begin{eqnarray}
&&C_v=T\frac{\partial S}{\partial T}\Big|_v\propto \left(-\frac{T-T_c}{T_c}\right)^{-\alpha},\nonumber\\
&&\eta=\frac{v_s-v_l}{v_c}\propto \left(-\frac{T-T_c}{T_c}\right)^{\beta},\nonumber\\
&&\kappa_T=-\frac{1}{v}\frac{\partial v}{\partial P}\Big|_T\propto \left(-\frac{T
-T_c}{T_c}\right)^{-\gamma},\nonumber\\
&& P-P_c \propto (v-v_c)^{\delta},\label{eq:27a}
\end{eqnarray}
where $``c"$ denotes the quantity  at the
critical point of the system.

In order to compute the critical exponent
$\alpha$ for $D=5$, we rewrite the entropy of
black hole (Eq.~(\ref{eq:7a})) as
$S=\frac{27\pi^2v^3}{128}\left(1+\frac{32\tilde{\alpha}_{GB}}{3v^2}\right)$.
Obviously this entropy $S$ is independent of $T$
for the constant value of specific volume $v$, so
we conclude that the critical exponent
$\alpha=0$. To obtain the other exponents, we introduce the expansion parameters
\begin{eqnarray}
\tau=t+1, \quad \nu=\omega+1,\label{eq:29a}
\end{eqnarray}
and expand this equation of state (\ref{eq:21e}) near the critical point
\begin{eqnarray}
p=1+a_{10}t+a_{11}t\omega+a_{03}\omega^3+\mathcal{O}(t\epsilon^2,\epsilon^4).\label{eq:30a}
\end{eqnarray}
During the phase transition, the pressure remains
constant
\begin{eqnarray}
&&p=1+a_{10}t+a_{11}t\omega_s+a_{03}\omega_s^3=1+a_{10}t+a_{11}t\omega_l+a_{03}\omega_l^3, \nonumber\\
\Rightarrow && a_{11}t\left(\omega_s-\omega_l\right)+a_{03}\left(\omega_s^3-\omega_l^3\right)=0,\label{eq:32a}
\end{eqnarray}
where $\omega_s$ and $\omega_l$ denote the `volume' of small and large black holes.

Using Maxwell's area law, we obtain
\begin{eqnarray}
\int^{\omega_s}_{\omega_l}\omega\frac{dp}{d\omega}d\omega=0 \Rightarrow a_{11}t\left(\omega_s^2-\omega_l^2\right)
+\frac{3}{2}a_{03}\left(\omega_s^4-\omega_l^4\right)=0.\label{eq:33a}
\end{eqnarray}
With Eqs.~(\ref{eq:32a})(\ref{eq:33a}), the
nontrivial solutions appear only when
$a_{11}a_{03}t<0$. Then we can get
\begin{eqnarray}
\omega_s=\frac{\sqrt{-a_{11}a_{03}t}}{3|a_{03}|}\approx3.23\sqrt{-t}, \quad
\omega_l=-\frac{\sqrt{-a_{11}a_{03}t}}{3|a_{03}|}\approx-3.23\sqrt{-t}.\label{eq:34a}
\end{eqnarray}
Therefore, we have
\begin{eqnarray}
\eta=\omega_s-\omega_l=2\omega_s=6.46\sqrt{-t}\Rightarrow \beta=1/2.\label{eq:35a}
\end{eqnarray}

The isothermal compressibility can be computed as
\begin{eqnarray}
\kappa_T=-\frac{1}{v}\frac{\partial v}{\partial P}\Big|_{v_c}\propto
-\frac{1}{\frac{\partial p}{\partial \omega}}\Big|_{\omega=0}=\frac{2}{3t},\label{eq:36a}
\end{eqnarray}
which indicates that the critical exponent $\gamma=1$. Moreover, the shape of the critical
isotherm $t=0$ is given by
\begin{eqnarray}
p-1=-\omega^3\Rightarrow \delta=3.\label{eq:37a}
\end{eqnarray}

Evidently in the grand canonical ensemble, these
critical exponents of the five-dimensional
spherical charged GB-AdS black holes coincide
with those of the Van der Waals liquid-gas system
\cite{Kubiznak:2012wp}.

\section{Phase transition at the critical point and Ehrenfest's equations}
\label{4s}

For Van der Waals liquid-gas system, the
liquid-gas structure does not change suddenly but
undergoes the second order phase transition at
the critical point $(V=V_c, T=T_c, P=P_c)$. This
is described by the Ehrenfest's description
\cite{Linder,Stanley}. In conventional
thermodynamics, Ehrenfest's description consists
of the first and second Ehrenfest's equations
\cite{sNieuwenhuizen,Zemansky}
\begin{eqnarray}
&&\frac{\partial P}{\partial T}\Big|_S=\frac{C_{P2}-C_{P1}}{TV(\zeta_2-\zeta_1)}
=\frac{\Delta C_P}{TV\Delta\zeta},\label{eq:38a}\\
&&\frac{\partial P}{\partial T}\Big|_V=\frac{\zeta_2-\zeta_1}{\kappa_{T2}-\kappa_{T1}}
=\frac{\Delta\zeta}{\Delta\kappa_{T}}.\label{eq:39a}
\end{eqnarray}
For a genuine second order phase transition, both of these equations have to be satisfied simultaneously.
Here $\zeta$ and $\kappa_T$ denote the volume expansion and isothermal compressibility coefficients
of the system respectively
\begin{eqnarray}
\zeta=\frac{1}{V}\frac{\partial V}{\partial T}\Big|_P,\quad
\kappa_T=-\frac{1}{V}\frac{\partial V}{\partial P}\Big|_T. \label{eq:40a}
\end{eqnarray}

Let us concentrate on the five-dimensional
spherical charged GB-AdS black hole in the grand
canonical ensemble. From Eq.~(\ref{eq:40a}), we
obtain
\begin{eqnarray}
V\zeta=\frac{\partial V}{\partial T}\Big|_P
=\frac{\partial V}{\partial S}\Big|_P\times\frac{\partial S}{\partial T}\Big|_P
=\frac{\partial V}{\partial S}\Big|_P\times\frac{C_P}{T}.\label{eq:41a}
\end{eqnarray}
The right hand side of Eq.~(\ref{eq:38a}) can be
expressed into
\begin{eqnarray}
\frac{\Delta C_P}{TV\Delta \zeta}=\left[\frac{\partial S}{\partial V}\Big|_P\right]_{r_+=r_c}
=\frac{3(r_c^2+2\tilde{\alpha}_{GB})}{4r_c^3},\label{eq:42a}
\end{eqnarray}
where the thermodynamic volume $V$ is described
in Eq.~(\ref{eq:12a}) and the subscript denotes
the physical quantities at the critical point. From Eq.~(\ref{eq:14a}),
the left hand side of Eq.~(\ref{eq:38a}) at the
critical point can be got as
\begin{eqnarray}
\left[\frac{\partial P}{\partial T}\Big|_S\right]_{r_+=r_c}
=\frac{3(r_c^2+2\tilde{\alpha}_{GB})}{4r_c^3}.\label{eq:43a}
\end{eqnarray}
Therefore, the first of Ehrenfest's equations can
be satisfied at the critical point.

Now let's examine the second of Ehrenfest's
equations. In order to compute $\kappa_T$, we
make use of the thermodynamic identify
\begin{eqnarray}
\frac{\partial V}{\partial P}\Big|_T\times\frac{\partial P}{\partial T}\Big|_V
\times\frac{\partial T}{\partial V}\Big|_P=-1.\label{eq:44a}
\end{eqnarray}
Considering Eq.~(\ref{eq:40a}), we can have
\begin{eqnarray}
\kappa_T V=-\frac{\partial V}{\partial P}\Big|_T=\frac{\partial T}{\partial P}\Big|_V
\times\frac{\partial V}{\partial T}\Big|_P
=\frac{\partial T}{\partial P}\Big|_V V\zeta.\label{eq:45a}
\end{eqnarray}
which reveals the validity of the second
Ehrenfest equations at the critical point.
Moreover, the right hand side of
Eq.~(\ref{eq:39a}) is given by
\begin{eqnarray}
\frac{\Delta\zeta}{\Delta\kappa_T}=\left[\frac{\partial P}{\partial T}\Big|_V\right]_{r_+=r_c}
=\frac{3(r_c^2+2\tilde{\alpha}_{GB})}{4r_c^3}.\label{eq:46a}
\end{eqnarray}
Using Eqs.~(\ref{eq:42a}) and (\ref{eq:46a}), the
Prigogine-Defay (PD) ratio $(\Pi)$ \cite{Nieuwenhuizen} is
\begin{eqnarray}
\Pi=\frac{\Delta C_P\Delta \kappa_T}{Tv(\Delta\zeta)^2}=1.\label{eq:47a}
\end{eqnarray}
Hence in the grand canonical ensemble this phase
transition at the critical point in the
five-dimensional charged GB-AdS black hole is of
the second order. This result is consistent with
the nature of the liquid-gas phase transition at
the critical point.

\section{Closing remarks}
\label{5s}

In this paper we have studied the thermodynamic
behaviors in the grand canonical ensemble by
fixing the potential of D-dimensional charged
GB-AdS black holes in an extended phase space by
treating the cosmological constant and their
conjugate quantity as thermodynamic variables. We
have written out the equations of state and
examined the phase structures by using the
standard thermodynamic techniques. We have shown
that only five-dimensional spherical GB-AdS black
holes admit a first order small-large black hole
phase transition when its potential is fixed
within the range $0<\Phi<\frac{\sqrt{3}\pi}{4}$,
which resembles the liquid-gas phase transition
in fluids. For the other higher dimensional and
topological charged GB-AdS black holes, we have
not found the phase transition. When we take the
limiting case with vanishing GB parameter, we
conclude that in the grand canonical ensemble,
there is no criticality for the RN-AdS black
holes in the extended phase space. This result is
independent of the spacetime dimensions and
topologies.

We have also computed the critical exponents of
the phase transition and found that in the fixed
potential ensemble the thermodynamic exponents
associated with the five-dimensional spherical
charged GB-AdS black hole coincide with those of
the Van der Waals fluid. We finally analyzed the
phase transition at the critical point by
employing Ehrenfest's scheme and verified that
both of the Ehrenfest's equations can hold at the
critical point. This shows that in resemblance
with the liquid-gas phase transition, at the
critical point the phase transition of the
five-dimensional charged GB-AdS black hole in the
fixed potential ensemble is of the second order.

{\bf Acknowledgments}

This work was supported by the National Natural Science Foundation of China.
D.C.Z are extremely grateful to Wei Xu for useful discussions.

\end{document}